# The Radius, Composition, Albedo, and Absolute Magnitude of Planet Nine Based on Exoplanets with Teq ≤ 600 K and the Planet Nine Reference Population 3.0

David G. Russell and Terry L. White

**ABSTRACT**

Evidence suggests the existence of a large planet in the outer Solar System, Planet Nine, with a predicted mass of 6.6 +2.6/-1.7 M⊕ (Brown et al., 2024). Based on mass-radius composition models, planet formation theory, and confirmed exoplanets with low mass and radius uncertainty and equilibrium temperature less than 600 K, we determine the most likely composition for Planet Nine is a mini-Neptune with a radius in the range 2.0 to 2.6 R⊕ and a H-He envelope fraction in the range of 0.6% to 3.5% by mass. Using albedo estimates for a mini-Neptune extrapolated from V-band data for the Solar System's giant planets gives albedo values for Planet Nine in the range of 0.47 to 0.33. Using the most likely orbit and aphelion estimates from the Planet Nine Reference Population 3.0, we estimate Planet Nine's absolute magnitude in the range of - 6.1 to - 5.2 and apparent magnitude in the range of + 21.9 to 22.7. Finally, we estimate that, if the hypothetical Planet Nine exists and is detected by upcoming surveys, it will have a resolvable disk using some higher resolution world-class telescopes.

**1. Introduction**

The Planet Nine Hypothesis (PNH) (Batygin et al., 2019) was developed to explain a large number of observational oddities present in the outer Solar System. Chief among these oddities is the clustering in the longitude of perihelion, $\varpi$, of stable, detached, extreme trans-Neptunian objects (ETNOs), which can be explained by the secular gravitational shepherding of a hypothetical distant giant planet, a Planet Nine, orbiting in the outer regions of our Solar System on a moderately-inclined, eccentric orbit. In addition, the PNH can explain the orbital detachment from Neptune of these same ETNOs and also explain the large number of highly-inclined or retrograde ETNOs and Centaurs. Finally, the observed ETNO perihelion gap between 50 and 65 au (Oldroyd and Trujillo, 2021); the inward scattering of detached Inner Oort Cloud objects to populate the Kuiper Belt and the inner Solar System (Batygin and Brown, 2021); the observed erosion of Neptune's higher order n:1 resonant populations (Clement and Sheppard, 2021), (Porter et al., 2024) and the many observed low-inclination Centaurs with semi-major axes ≳ 100 au that are broadly distributed in perihelion space (Batygin et al., 2024) can also be easily explained by a Planet Nine.

The PNH uses a secular gravitation model of Planet Nine to relate the observed clustering of detached ETNOs to Planet Nine's orbital shape and orientation. The statistical sampling of thousands of N-body simulations using the PNH secular gravitational model predicts the most likely probability distributions of Planet Nine's: mass, $M_9$; semi-major axis, $a_9$; inclination, $i_9$; eccentricity, $e_9$; ascending node, $\Omega_9$; argument of perihelion, $\omega_9$ and longitude of perihelion, $\varpi_9$, where $\varpi_9 = \omega_9 + \Omega_9$. The PNH says nothing about Planet Nine's true anomaly, radius, albedo or composition beyond its mass, so that leaves a lot of room for speculation about its orbital location and makeup. But clues to the makeup of Planet Nine lie in its evolution. One extreme says that Planet Nine was formed in-situ as a fifth or sixth giant planet and perturbed early in its formation by one or more of the giant planets before it had much time to accrete a hydrogen/helium (H-He) atmosphere. In this case Planet Nine could be a failed-ice-giant-core and resemble the larger ETNOs. Most ETNOs are quite dark and reddish with very low albedos (0.02 - 0.1) which would make Planet Nine harder to detect. On the other hand, if Planet Nine was perturbed in-situ later in its formation and had time to form a mini-Neptune, then it would be much brighter and bigger and the albedo would be dominated by Rayleigh scattering (0.3 - 0.75) which would make it easier to detect.

One of the reasons the Planet Nine Reference Population (PNRP) (Brown, 2023) was developed was to specifically address the uncertainties in planetary radius and albedo to aid in calculating the apparent magnitude of Planet Nine. The PNRP consists of 100,000 Planets Nine (Brown et al., 2021) with the following parameters: $a_9$, $i_9$, $e_9$, $\varpi_9$ and $\Omega_9$ probability distributions and covariances from the results of thousands of N-body simulation runs; a true anomaly picked for every degree between 0º and 360º; albedos are selected from 0.2 (½ Neptune's albedo) to 0.75 (a Rayleigh-scattering albedo); masses from 3 to 20 $M_⊕$, and finally radii from 1 to 6.7 $R_⊕$ (a mini-Neptune). From the above values, the 100,000 synthetic Planets Nine are created and injected into each survey's data and each one must be detected multiple times in each survey to measure survey completeness and to be removed from the PNRP. The PNRP is currently in version 3.0 after the following Planet Nine survey searches eliminated various subpopulations from the full PNRP: the Zwicky Transient Facility Survey eliminated 56% of the original PNRP (Brown and Batygin, 2022); the Dark Energy Survey eliminated an additional 5% of the remaining PNRP (Belyakov et al., 2022); the Pan-STARRS1 Survey eliminated an additional 17% of the remaining PNRP, (Brown et al., 2024).

D. G. Russell et al.In this paper we constrain the most likely ranges for the radius, composition, absolute magnitude, and apparent magnitude of Planet 9 based on: the most recent mass estimate for Planet 9; the population of confirmed exoplanets with low mass and radius uncertainty and equilibrium temperature < 600 K; mass-radius composition models; findings derived from core-accretion planet formation theory; albedo estimates drawn from the Solar System giant planets, and the PNRP.

## 2. Possible Compositions for Planet Nine

In this section we aim to further narrow down the composition of Planet Nine. After the latest published survey using the Pan-STARRS1 archive (Brown et al., 2024), they estimate the remaining 22% of the PNRP has the median, 15.8 and 84.1 percentile values for $M_9$ as $M_9 = 6.6 + 2.6 / - 1.7\ M_\oplus$. The two most plausible composition classes for Planet Nine are:

a. A solid-rock-ice terrestrial composition. Mass-radius composition models can be used to predict the Planet Nine radius, $R_9$, for terrestrial compositions using models from Zeng et al., 2016 and Zeng et al., 2019. The predicted radii for Planet Nine masses of 4.9 $M_\oplus$, 6.2 $M_\oplus$, and 9.2 $M_\oplus$ are presented in Table 1 below. In addition to the more likely mixed rock-ice compositions for a planet formed in the outer Solar System, the table includes the radii for an Earth-like pure rock terrestrial planet with a 30%-by-mass iron core.

**Table 1.**
Planet Nine Planet Nine Terrestrial Radii versus Composition and Mass Assumptions

| Planet Nine Terrestrial Composition (%/%) | $R_9$ for Various $M_9$ Assumptions | | |
|---|---|---|---|
| | $M_9 = 4.9\ M_\oplus$ | $M_9 = 6.6\ M_\oplus$ | $M_9 = 9.2\ M_\oplus$ |
| | ($R_\oplus$) | ($R_\oplus$) | ($R_\oplus$) |
| 30 Fe / 70 SiO$_2$ | 1.55 | 1.66 | 1.81 |
| 30 H$_2$O / 70 SiO$_2$ | 1.86 | 2.00 | 2.18 |
| 50 H$_2$O / 50 SiO$_2$ | 1.95 | 2.10 | 2.29 |
| 70 H$_2$O / 30 SiO$_2$ | 2.04 | 2.20 | 2.39 |

b. A mini-ice-giant composition or "mini-Neptune." Table 2 provides predicted radii for possible mini-Neptune compositions with H-He mass fractions ranging from 0.1% to 10%. Radius values are interpolated to an age of approximately 4.5 Gyr from the Solar metallicity 1 Gyr and 10 Gyr tables in Lopez and Fortney, 2014 for a stellar flux, $F$, of 0.1 $F_\oplus$, where $F_\oplus$ is the Earth's stellar flux.

**Table 2.**

Planet Nine Mini-Neptune Radii versus Composition and Mass Assumptions

| Planet Nine Mini-Neptune Mass Fraction Composition of H-He (%) | $R_9$ for Various $M_9$ Assumptions | | |
|---|---|---|---|
| | $M_9 = 4.9\ M_\oplus$ | $M_9 = 6.6\ M_\oplus$ | $M_9 = 9.2\ M_\oplus$ |
| | ($R_\oplus$) | ($R_\oplus$) | ($R_\oplus$) |
| 0.1 | 1.62 | 1.73 | 1.86 |
| 0.5 | 1.85 | 1.96 | 2.08 |
| 1.0 | 2.00 | 2.11 | 2.23 |
| 2.0 | 2.21 | 2.32 | 2.43 |
| 5.0 | 2.69 | 2.80 | 2.90 |
| 10.0 | 3.35 | 3.45 | 3.53 |

The radii in Tables 1 and 2 indicate that within the mass range 4.9 to 9.2 $M_\oplus$, regardless of composition, the radius of Planet Nine should exceed 1.6 $R_\oplus$. The maximum radius for Planet Nine, based upon realistic possible compositions and assuming a maximum mass of 9.2 $M_\oplus$, would be 2.39 $R_\oplus$ for a 70% water 30% SiO$_2$ rock-ice terrestrial composition or 3.53 $R_\oplus$ for a 10% H-He mini-Neptune composition. Despite the large range of possible radii and compositions for Planet Nine presented in Tables 1 and 2, the results of planet formation models and the confirmed exoplanet population can be used to identify a narrower range of likely radii for Planet Nine.

3. **Composition Suggested from the Core Accretion Scenario of Planet Formation**

Core accretion is the primary mechanism believed to be responsible for the formation of giant planets up to a few Jupiter masses (Pollack et al., 1996 and Schlaufman, 2018). In the core accretion model, a solid rock or rock-ice core grows through planetesimal and pebble accretion. Once a critical mass is reached the core begins to accrete substantial amounts of H-He. As the growing planet reaches the crossover mass where the core and H-He envelope are approximately of equal mass, the planet can experience runaway gas accretion (Pollack et al., 1996; see recent review by Raymond et al., 2020). Studies have consistently found that the critical core mass above which a core begins to accrete substantial amounts of H-He is approximately 4 $M_\oplus$ (Ida & Lin, 2004; Hasegawa, 2016 and Alessi et al., 2020). Furthermore, Hasegawa, (2016) found that all planets exceeding the critical core mass for gas accretion should have both solids and a gaseous atmosphere. Finally, Alessi et al., (2020) found that approximately 2 to 3 $M_\oplus$ is the transition between planets with radii predominantly determined from the solid cores and those with larger masses that have a large contribution to the radius from a gaseous envelope. Brown et al. (2024) estimated the mass of Planet Nine to be 6.6 + 2.6 / - 1.7 $M_\oplus$ which is well above the critical mass needed to begin accretion of a substantial H-He envelope, though not massive enough to initiate runaway accretion. It is therefore expected that Planet Nine is not a solid rock-ice terrestrial body (Table 1), but is instead a "mini-Neptune" with a rock-ice core and an outer H-He envelope (Table 2).

4. **The Probable Radius Range of Planet Nine from the Exoplanet Population**

In this section we aim to further narrow down the radius range of Planet Nine. The catalog of confirmed exoplanets in the NASA Exoplanet Archive, described by Akeson et al., (2013), and Christiansen et al., (2025) is nearing 6000 as of July 2025. One difficulty with the current exoplanet sample is that the detection of short orbital period planets with high equilibrium temperatures, $T_{eq}$, is favored over low equilibrium temperature planets such as those in the outer Solar System. When considering exoplanets with an H-He atmosphere, short orbital period planets can have significant radius inflation from both high stellar flux, see Lopez and Fortney, (2014), and tidal inflation, see Millholland et al., (2020). For example, an 8.5 $M_\oplus$ exoplanet with a 5% H-He envelope by mass have predicted radii of 3.31 $R_\oplus$ and 2.72 $R_\oplus$ for stellar flux values of 1000 $F_\oplus$ and 0.1 $F_\oplus$ respectively Lopez and Fortney, (2014). Millholland et al., (2020) found that sub-Saturn radius exoplanets with shorter orbital periods could experience significant tidal inflation which leads to overestimated H-He fractions if tidal inflation is ignored. In addition, with sufficient stellar flux, mini-Neptune's can be stripped of their H-He envelope from photoevaporation, leaving behind the core as a high mass terrestrial composition exoplanet or "hot earth" Dai et al., (2019). For each of these reasons, high equilibrium temperature exoplanets ($T_{eq}$ > 700 K) provide poor analogs for determining the likely radius and composition of Planet Nine.

In order to evaluate the likely radius range and composition for Planet Nine, exoplanets with equilibrium temperature less than 600 K were selected from the NASA exoplanet archive. To be included in the sample, the exoplanets were required to meet the following criteria: a mass uncertainty less than + / - 27%; a radius uncertainty less than + / - 8.5%; a stellar flux less than 25 $F_\oplus$ or $T_{eq}$ < 600 K and a mass in the range of 4.9 to 9.2 $M_\oplus$. These selection criteria result in a total sample of 27 exoplanets from the catalog of exoplanets with $T_{eq} \leq 600$ K maintained by Russell, (2023). The full $T_{eq} \leq 600$ K catalog drawn from the NASA Exoplanet Archive as of July 2025, contains 142 total exoplanets meeting the mass uncertainty, radius uncertainty, and $F_\oplus/T_{eq}$ selection criteria. The strict requirements for mass and radius uncertainty ensure better accuracy of possible compositions when applying mass-radius-composition models (e.g. Otegi et al., (2020) and Parc et al., (2024)). It's important to note that the exoplanets in the $T_{eq} \leq 600$ K sample have radii that will be significantly less affected by photoevaporation and tidal inflationary effects than higher equilibrium temperature planets. This can be seen by looking at the data tables and analysis of Lopez and Fortney, (2014) and Millholland et al., (2020).

One important characteristic of the full sample (N = 142) of $T_{eq} \leq 600$ K planets is that there are no terrestrial composition exoplanets with a mass exceeding 2.9 $M_\oplus$. This is consistent with the predictions of the core accretion planet formation scenario which predicts exoplanets with a mass exceeding approximately 4 $M_\oplus$ will accrete a substantial H-He envelope (Ida & Lin, 2004; Hasegawa, 2016 and Alessi et al., 2020). This also supports the prediction that Planet Nine is not a rock-ice terrestrial planet, but is instead a mini-Neptune. Finally, the 2.9 $M_\oplus$ upper mass limit for terrestrial composition planets in the $T_{eq} \leq 600$ K catalog also supports the expectation that the $T_{eq} \leq 600$ K sample is less affected by radius inflation and photoevaporation of H-He envelopes. The higher equilibrium temperature exoplanet population includes numerous massive terrestrial composition exoplanets that are most likely the stripped cores of Neptunes and mini-Neptunes Dai et al., (2019). The sample of exoplanets meeting the selection criteria described above are listed in Table 3 below.

**Table 3**
Exoplanets with $T_{eq} \leq 600$ °K and Mass in the Range of 4.9 to 9.2 $M_\oplus$

| Exoplanet Name | Period (days) | Mass ($M_\oplus$) | Radius ($R_\oplus$) | $F / T_{eq}$ ($F_\oplus$ / °K) | References* |
|---|---|---|---|---|---|
| K2-3 b | 10.05 | 5.11 + 0.65 / - 0.64 | 2.078 + 0.076 / - 0.067 | 10.5 / 501 | Diamond-Lowe et al., 202 |
| LHS 1140 b | 24.74 | 5.60 +/- 0.19 | 1.730 +/- 0.025 | 0.43 / 226 | Cadieux et al., 2024 |
| TOI-1801 b | 10.64 | 5.74 +/- 1.46 | 2.08 +/- 0.12 | - / 393-494 | Mallorquín et al., 2023 |
| K2-146 b | 2.64 | 5.77 +/- 0.18 | 2.05 +/- 0.06 | 20.7 / 534 | Hamann et al., 2019 |
| TOI-406 c | 13.18 | 6.57 + 1.00 / - 0.90 | 2.08 + 0.16 / - 0.15 | - / 368 | Lacadelli et al., 2024 |
| TOI-1468 c | 15.53 | 6.64 +/- 0.68 | 2.064 +/- 0.044 | 2.15 / 338 | Chaturvedi et al., 2022 |
| K2-146 c | 4.00 | 7.49 +/- 0.24 | 2.19 +/- 0.07 | - / < 534 | Hamann et al., 2019 |
| HD 23472 b | 17.67 | 8.32 + 0.78 / - 0.79 | 2.00 + 0.11 / - 0.10 | 16.0 / 543 | Barros et al., 2022 |
| TOI-2443 b | 15.67 | 4.88 +/- 1.1 | 2.363 +/- 0.066 | - / 519 | Naponiello et al., 2025 |
| Kepler 289 d | 66.03 | 5.33 + 0.43 / - 0.42 | 3.03 +/- 0.08 | ? | Greklek-McKeon et al., 20 |
| TOI-4438 b | 7.45 | 5.4 +/- 1.1 | 2.52 +/- 0.13 | 6.00 / 435 | Goffo et al., 2024 |
| Kepler 51 d | 130.18 | 5.70 +/- 1.12 | 9.46 +/- 0.16 | 2.466* / - | Libby-Roberts et al., 2020 |
| TOI-270 c | 5.66 | 6.15 +/- 0.37 | 2.355 +/- 0.064 | - / 488 | Van Eylen et al., 2021 |
| Kepler 305 d | 16.74 | 6.20 + 1.76 / - 1.34 | 2.76 +/- 0.12 | 25.52* / - | Leleu et al., 2023 |
| Kepler 79 e | 81.06 | 6.3 +/- 1.0 | 3.414 +/- 0.129 | 18.78* / - | Yofee et al., 2021 |
| Kepler 87 c | 191.23 | 6.4 +/- 0.8 | 6.14 +/- 0.29 | - / 403 | Ofir et al., 2014 |
| LP 791-18 c | 4.99 | 7.1 +/- 0.7 | 2.438 +/- 0.096 | 2.64 / 324 | Peterson et al., 2023 |
| Kepler 26 c | 17.25 | 7.48 + 0.49 / - 0.48 | 3.11 +/- 0.14 | 6.63* / - | Leleu et al., 2023 |
| TOI-178 f | 15.23 | 7.72 + 1.67 / - 1.62 | 2.287 +/- 0.11 | - / 521 | Leleu et al., 2021 |
| HD 191939 c | 28.58 | 8.0 +/- 0.1 | 3.195 +/- 0.075 | 21.0 / 600 | Orell-Miquel et al., 2023 |
| TOI-1803 c | 12.89 | 8 +1 / - 2 | 4.29 +/- 0.08 | 22 / 588 | Zingales et al., 2025 |
| LTT 3780 c | 12.25 | 8.04 + 0.50 / - 0.48 | 2.39 + 0.10 / - 0.11 | 2.76 / 359 | Bonfanti et al., 2024 |
| Kepler 49 c | 10.91 | 8.38 + 0.92 / - 0.89 | 2.444 + 0.083 / - 0.082 | 12.34* / - | Leleu et al., 2023 |
| GJ 1214 b | 1.58 | 8.41 + 0.36 / - 0.35 | 2.733 + 0.033 / - 0.031 | 17.2 / 567 | Mahajan et al. 2024 |
| K2-18 b | 32.94 | 8.63 +/- 1.35 | 2.610 +/- 0.087 | 1.005 / 255 | Benneke et al., 2019 |
| TOI-269 b | 3.70 | 8.80 +/- 1.40 | 2.77 +/- 0.12 | 19.0 / 531 | Cointepas et al., 2021 |
| HD 136352 d | 107.25 | 8.82 +0.93 / - 0.92 | 2.562 + 0.088 / - 0.079 | 5.74 / 431 | Delraz et al., 2021 |

\* - References for each exoplanet can also be found by searching the exoplanet identification in the NASA Exoplanet Archive.

The first 8 exoplanets in Table 3 can be modeled with an H-He mass fraction < 1% by mass using the tables of Lopez and Fortney, (2014). The remaining 19 planets in Table 3 have mass-radius values consistent with a H-He mass fraction ranging from 1% by mass to > 20% by mass. Of this second group, 16 out of 19 are consistent with H-He mass fractions in the range 1 to 9% by mass according to the models of Lopez and Fortney, (2014), whereas the remaining 3 are larger than Uranus and Neptune including 2 very large diameter exoplanets characterized as "Super-puffs," (e.g. Libby-Roberts et al., 2020). In order to assess the most likely value for the radius of Planet Nine, three median values were calculated for the $T_{eq} < 600$ K sample in Table 3 and are: a median radius of 2.07 $R_\oplus$ for exoplanets in Table 3 with a H-He fraction < 1%; a median radius of 2.44 $R_\oplus$ for the full sample of exoplanets in Table 3; and a median radius of 2.73 $R_\oplus$ for the exoplanets in Table 3 with a H-He fraction > 1%.

The < 1% H-He group sets a likely lower radius limit for Planet Nine, whereas the H-He > 1% group sets the likely upper radius limit, assuming Planet Nine has a composition and radius consistent with the $T_{eq} < 600$ K exoplanet population discovered to date in the mass range 4.9 to 9.2 $M_\oplus$. As can be seen in Table 3, the stellar flux of the exoplanets, while relatively low at roughly 10 $F_\oplus$, is still substantially higher than the expected solar flux in the outer Solar System. Therefore, these median radii are corrected from 10 $F_\oplus$ to 0.1 $F_\oplus$ using the Tables of Lopez and Fortney, (2014) and Table 2 in this analysis. The resulting median radii values and H-He fractions are: 1.98 $R_\oplus$ → 0.6% H-He by mass; 2.31 $R_\oplus$ → 2 % H-He by mass and 2.57 $R_\oplus$ → 3.5% H-He by mass.

Based upon the $T_{eq} < 600$ K exoplanet sample, the most likely radius for Planet Nine falls in the range 2.0 to 2.6 $R_\oplus$ and the most likely composition is a mini-Neptune with an H-He fraction in the range 0.6 to 3.5 % by mass. One caveat to the application of the Lopez and Fortney, (2014) models is that they are based upon an H-He envelope with an Earth-like composition core. However, Alessi et al., (2020) found that radii of exoplanets massive enough to accrete an H-He envelope will be within 5% even if comparing an Earth-like composition core to an exoplanet with a 33% ice fraction. The effect of the exact core composition is therefore expected to have no more than a 0.10 to 0.13 $R_\oplus$ impact on the predicted radius range for Planet Nine, derived from this exoplanet sample.

5. **The V-band Albedo of Planet Nine**

   Two methods were used to estimate the V-band albedo, $A_9$, of Planet Nine as a function of heliocentric distance, $r$.

   a. The first method used the mean V-band albedo data from Mallama et al., (2016) only for Uranus and Neptune to represent the most likely value for $A_9$ at 0.465, which is used to calculate the absolute magnitude for Planet Nine and is shown in column 2 of Table 4 below, specifically for radius values of 1.98, 2.31, and 2.57 $R_\oplus$.

   b. The second method used to estimate the albedo of Planet Nine was to simply extrapolate a power-law curve fit of the V-band albedo of all the giant planets using the same data from Mallama et al., (2016), which gives a very good fit with only two parameters given by and is shown in Fig. 1 below and explains over 91% of the variance in the albedo data.

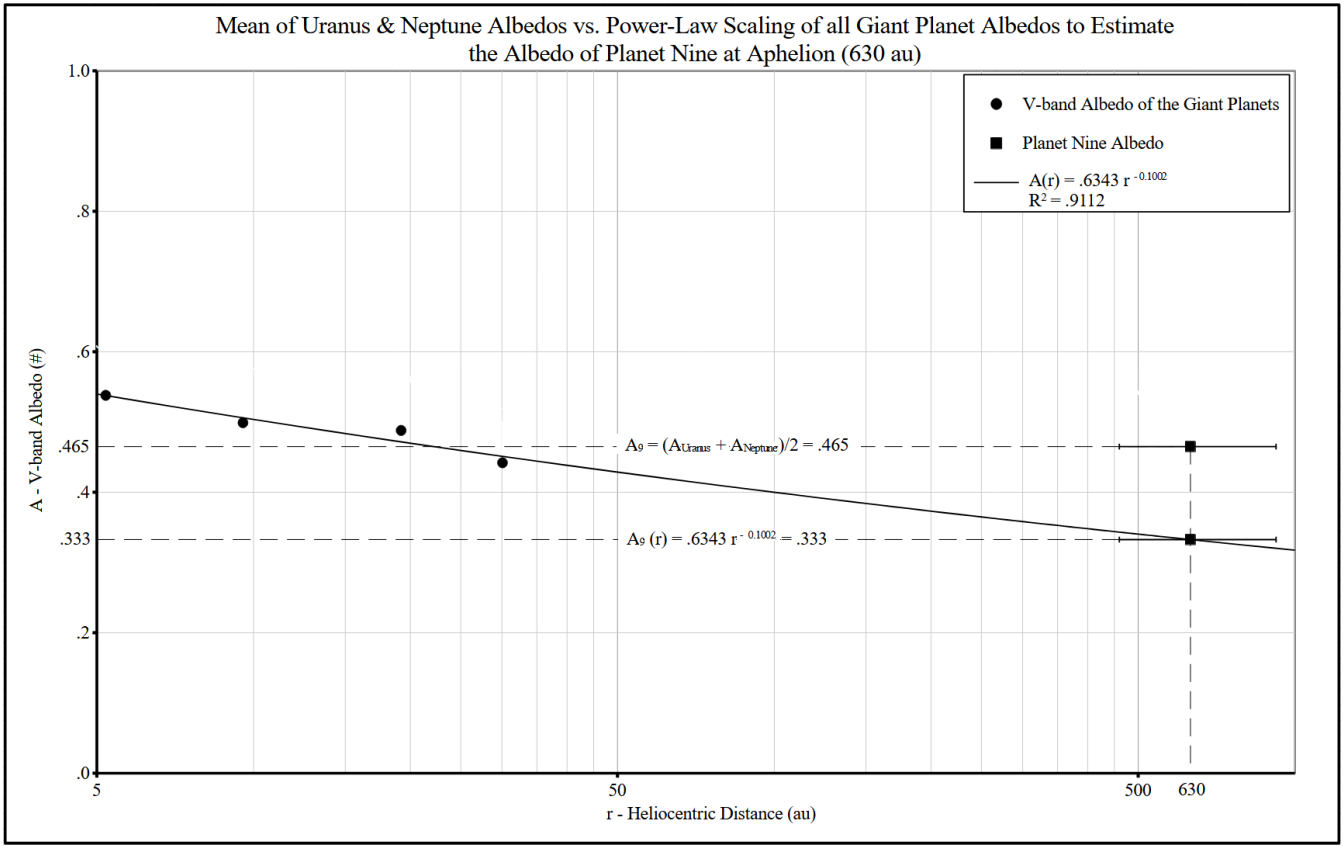

**Figure 1.** The Mean of Uranus and Neptune Albedos vs. Power Law Scaling of all giant Planet Albedos to Estimate the Albedo of Planet Nine at Aphelon.

Brown et al., (2024) report that the ZTF+DES+PS1 surveys have removed 78% of the PNRP, and the current distance of Planet Nine, $r_9$, is now 550 + 250 / - 180 au, based on the remaining 22% of the PNRP. They also report that the aphelion, $Q_9$, is now 630 + 290 / - 170 au. Aphelion, however, is the most likely location for any body on an eccentric orbit simply due to residence time considerations. We therefore use the aphelion location of 630 au, rather than the current distance, no matter what the remaining 34% of the PNRP favors, because the remaining PNRP includes the most likely aphelion location on the maximum likelihood orbit as shown as a white asterisk in Figure 2 below. The remaining PNRP is shown in a Mollweide equatorial projection centered at longitude of 180° and latitude of 0°. The approximate centroid of the PNRP maximum relative probability map is marked with a white cross and the relative probability key is at the bottom. The galactic center and anti-center are marked with blue crosses. The galactic limits of ± 15° in galactic latitude denote the galactic plane, where detections are difficult, and are shown as heavy black lines, as well as the ecliptic. Lighter black lines denote every 45° in equatorial longitude and latitude.

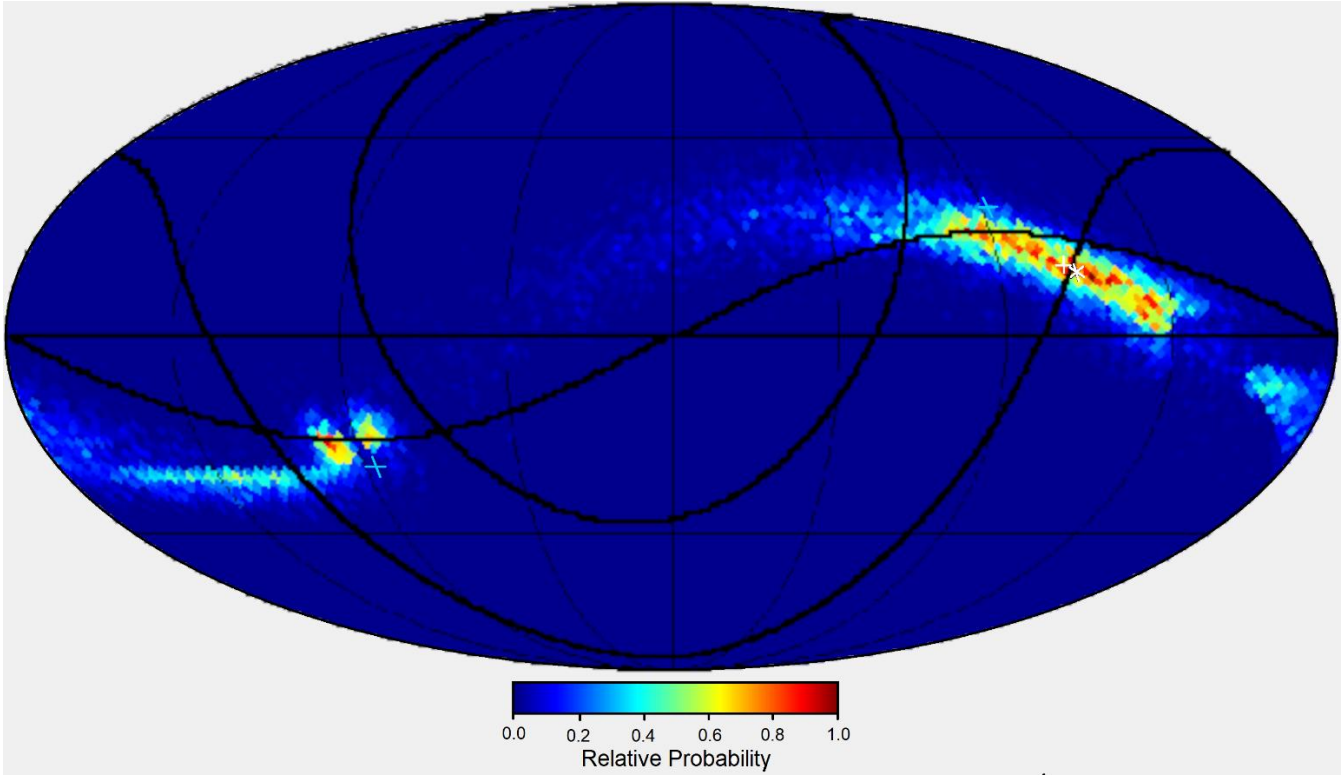

**Figure 2.** The remaining 24% of the PNRP after the ZTF+DES+PS1 surveys.

Using power law scaling $A_9 = 0.333$ and is shown in column 3 of Table 4 below. Therefore the most likely range for $A_9$ is from 0.465 to 0.333.

**Table 4.**

Planet Nine Absolute Magnitudes versus Mini-Neptune Median Radii and Albedo Assumptions

| $R_9$ | $H_9$ for various $A_9$ assumptions | |
|---|---|---|
| | $A_9 = 0.465$ | $A_9 = 0.333$ |
| ($R_\oplus$) | (#) | (#) |
| 1.98 | - 5.58 | - 5.20 |
| 2.31 | - 5.92 | - 5.53 |
| 2.57 | - 6.13 | - 5.77 |

## 6. The Absolute Magnitude and Angular Extent of Planet Nine at Aphelion

The absolute magnitude of Planet Nine, $H_9$, can be calculated for different radius and geometric albedo values and is calculated for various $R_9$ and $A_9$ assumptions in Table 4 above. The apparent magnitude of Planet Nine, $m_9$, is calculated for Planet Nine at aphelion (630 au) for the radii and albedo assumptions made in Table 4 and listed in Table 5 below. Also, the angular extent of Planet in milli-arc-seconds (mas) is shown in Table 5, column 4 below for $r = 630$ (aphelion).

**Table 5.**

Planet Nine Apparent Magnitudes and Angular Extent at Aphelion versus Mini-Neptune Median Radii and Albedo Assumptions

| $R_9$ | Planet Nine Apparent Magnitude $m_9$ @ Aphelion for two albedo $A_9$ assumptions | | Planet Nine Angular Extent at Aphelion |
|---|---|---|---|
| | $A_9 = 0.465$ | $A_9 = 0.333$ | |
| $(R_\oplus)$ | (#) | (#) | (mas) |
| 1.98 | +22.41 | +22.70 | 55.3 |
| 2.31 | +22.07 | +22.46 | 64.5 |
| 2.57 | +21.86 | +22.22 | 71.7 |

A quick survey of the resolution of some world-class telescopes is shown in Table 6 below.

**Table 6.**

Angular Resolution of World-Class Optical, Infrared and Radio Telescopes Compared to Planet Nine's Apparent Angular Diameter

| Telescope / Camera | Maximum Angular Pixel Scale Resolution (mas/pixel) | Band (nm/um/mm) | Resolvability of Planet Nine's Disk (requires 3 - 5 pixels) |
|---|---|---|---|
| Hubble / ACS - WFC | 50 | 370 → 1160 nm | No |
| Keck / NIRC2 - Narrow Camera | 10 | 900 → 5300 nm | Yes |
| JWST / NIRISS | 65 | 300 → 2000 um | No |
| ALMA (Band 10) | 20 | 0.3 → 0.4 mm | Yes |

From Table 5, Planet Nine has an angular extent of 55 to 72 mas, which is near the ultimate pixel scale resolution of Hubble and JWST but definitely resolvable with Keck and ALMA. In order to accurately measure a disk they will require at least 3 to 5 pixels across the disk to deconvolve the disk angular diameter from the finite resolution of the telescope. Keck and arrays like ALMA are all very capable at 10 and 20 mas, respectively.

**Conclusion**

Based upon the mass estimate of Planet Nine Brown et al., (2024), planet formation theory, composition models, and the population of lower equilibrium temperature exoplanets with low mass and radius uncertainty, the most likely composition for Planet Nine is a mini-Neptune with a radius in the range 2.0 to 2.6 $R_\oplus$ and with a H-He mass fraction in the range 0.6% to 3.5% by mass. From these predicted radius values, Planet Nine's absolute magnitude would be in the range of - 5.6 to - 6.2, assuming an albedo value similar to that of Uranus and Neptune, or -5.2 to -5.8 using an albedo of 0.33 from power-law scaling of all the giant planet's albedos from Mallama et al., (2016). If we assume Planet Nine is at the most likely aphelion distance of 630 au, on its maximum likelihood orbit, then Planet Nine's apparent magnitude, $m_9$, should be in the range of + 21.9 to + 22.7 for the range of radii and albedo assumptions made in this work. Finally the prospects for resolving Planet Nine's disk by Keck and ALMA are good. We expect that the PNRP will be updated to version 4.0 once the Suraru/HSC survey for Planet Nine by Brown's team is completed (More, et al. 2022) and we may revisit these estimates, if necessary.

**Author Contact:**

David G. Russell: dgrussellastro@gmail.com

Terry L. White: If0k0nle@gmail.com

**Declaration of Competing Interest**

The authors declare that they have no known competing financial interests or personal relationships that could have appeared to influence the work reported in this paper.

**Data Availability**

The mass, radius, and equilibrium temperatures for the sample of planets discussed in this note was acquired from the NASA Exoplanet Archive via https://exoplanetarchive.ipac.caltech.edu/. The Planet Nine Reference Population V 3.0 was acquired from Michael E. Brown via https://data.caltech.edu/records/8fjad-x7y61. All underlying data in the tables and figures can be provided by the authors upon request.

**Acknowledgements**

This research has made use of several sources: NASA's Exoplanet Archive at https://exoplanetarchive.ipac.caltech.edu/ plus the Planet Nine Reference Population V 3.0 at https://data.caltech.edu/records/8fjad-x7y61 both operated by the California Institute of Technology, the Astrophysics Data System Bibliographic Services at https://ui.adsabs.harvard.edu/ operated by NASA and finally the arXiv eprint service at https://arxiv.org/ operated by Cornell University.